\documentclass[a4paper,11pt]{article}
\usepackage{pos}
\usepackage{graphicx}
\usepackage{amsmath}

\usepackage[utf8]{inputenc}
\usepackage{newtxtext,newtxmath}
\usepackage[]{units}
\usepackage{enumitem}
\usepackage{booktabs}
\usepackage{url}
\usepackage{mathtools}
\usepackage{subfigure}
\usepackage{wrapfig}
\usepackage{comment}
\usepackage{soul}
\usepackage{lineno}
\usepackage{tabularx}
\DeclareUnicodeCharacter{2212}{-}

\title{Projected sensitivity to light WIMP-like particles of the BULLKID-DM experiment}

\author*[a,b]{Matteo Folcarelli}
\author[e]{A.~Acevedo-Rentería}
\author[i]{L.~E.~Ardila-Perez}
\author[f]{L.~Bandiera}
\author[g]{M.~Calvo}
\author[a,b]{M.~Cappelli}
\author[l]{R.~Caravita}
\author[d]{F.~Carillo}
\author[g]{U.~Chowdhury}
\author[i]{D.~Crovo}
\author[b]{A.~Cruciani}
\author[h]{A.~D’Addabbo}
\author[g,a,b]{D.~Delicato}
\author[c,d]{M.~De Lucia}
\author[b]{G.~Del Castello}
\author[a,b]{M.~del Gallo Roccagiovine}
\author[h]{F.~Ferraro}
\author[h]{S.~Fu}
\author[i]{R.~Gartmann}
\author[d]{M.~Grassi}
\author[m,f]{V.~Guidi}
\author[h]{D.~Helis}
\author[c,d]{T.~Lari}
\author[f]{L.~Malagutti}
\author[m,f]{A.~Mazzolari}
\author[g]{A.~Monfardini}
\author[i]{T.~Muscheid}
\author[c,d]{D.~Nicolò}
\author[c,d]{F.~Paolucci}
\author[b]{D.~Pasciuto}
\author[a,b]{L.~Pesce}
\author[d]{C.~Puglia}
\author[a,b]{D.~Quaranta}
\author[c,d]{C.~M.~A.~Roda}
\author[c,d]{S.~Roddaro}
\author[f]{M.~Romagnoni}
\author[c,d]{G.~Signorelli}
\author[i]{F.~Simon}
\author[d]{A.~Tartari}
\author[e]{E.~Vázquez-Jáuregui}
\author[a,b]{M.~Vignati}
\author[h,o]{K.~Zhao}

\affiliation[a]{Dipartimento di Fisica, Sapienza Università di Roma, P. le A. Moro 2, 00185 Roma, Italy}
\affiliation[b]{INFN Sezione di Roma, P.le A. Moro 2, 00185 Roma, Italy}
\affiliation[c]{Dipartimento di Fisica "Enrico Fermi", Università di Pisa, Largo Bruno Pontecorvo 3, 56127 Pisa, Italy}
\affiliation[d]{INFN Sezione di Pisa, Largo Bruno Pontecorvo 3, 56127 Pisa, Italy}
\affiliation[e]{Instituto de Física, Universidad Nacional Autónoma de México, A.P. 20-364, Ciudad de México 01000, México}
\affiliation[f]{INFN Sezione di Ferrara, Via Saragat 1, 44122 Ferrara, Italy}
\affiliation[g]{University Grenoble Alpes, CNRS, Grenoble INP, Institut Néel, 38000 Grenoble, France}
\affiliation[h]{INFN Laboratori Nazionali del Gran Sasso, 67100 Assergi (AQ), Italy}
\affiliation[i]{Institute for Data Processing and Electronics, Karlsruhe Institute of Technology, Hermann-von-Helmholtz-Platz 1 76344, Eggenstein-Leopoldshafen, Germany}
\affiliation[l]{INFN - TIFPA, Via Sommarive 14, 38123 Povo (Trento), Italy}
\affiliation[m]{Dipartimento di Fisica e Scienze della Terra, Università di Ferrara, Via Saragat 1, 44100 Ferrara, Italy}
\affiliation[n]{Dipartimento di Neuroscienze e Riabilitazione, Università di Ferrara, Via Luigi Borsari 46, 44121 Ferrara, Italy}
\affiliation[o]{Gran Sasso Science Institute, Viale Francesco Crispi, 7 Rectorate, Via Michele Iacobucci, 2, 67100 L'Aquila AQ, Italy}

\emailAdd{matteo.folcarelli@uniroma1.it}

\abstract{BULLKID-DM is an experiment designed for the direct searches of particle dark matter candidates with mass around 1 GeV, or below, and cross-section with nucleons smaller than $10^{-40}$ cm$^2$. The detector consists of a stack of diced silicon wafers, acting as arrays of particle absorbers, sensed by multiplexed Kinetic Inductance Detectors. The target will amount to 800 g subdivided in more than 2000 silicon dice, with the aim of controlling the background from natural radioactivity by creating a fully active structure and by applying fiducialization techniques. In this work we present the projected sensitivity of BULLKID-DM to light WIMP-like particles  considering also the other future experiments in the field.}

\FullConference{
}

\begin{document}
\maketitle
\section{Introduction}  
WIMP-like particles with GeV/c$^2$ mass, or less, are highly intriguing candidates for Dark Matter (DM) \cite{DM}, but the detection is difficult as it requires nuclear recoil detectors with energy thresholds of hundreds or tens of eV \cite{SuperCDMS-CPD,CRESST-III,Edelweiss}. Moreover a thorough exploration of interactions with cross-sections lower than $10^{−40}$ cm$^2$ demands an experiment with around a kilogram of active mass and zero background, a prospect that is challenging for existing experiments and their planned upgrades. BULLKID-DM aims to create such an experiment by introducing a new detector concept. The target mass will amount to a total mass of around $800$ g and will be composed of $2300$ silicon dice sensed by cryogenic phonon-mediated Kinetic Inductance Detectors (KIDs). With respect to other solid-state experiments in the field, the aim is to control the background from natural radioactivity by creating a fully active array of detectors without inert materials in its structure, and by performing a fiducialization of the target, resulting in a fiducial mass of approximately $600$ g. In this work we present the projected sensitivity of the future experiment to the Spin-Independent (SI) cross-section of WIMPs with nucleons as a function of their hypothetical mass.

\section{BULLKID-DM roadmap}
\begin{figure}[h]
    \centering
    \includegraphics[width=1\linewidth]{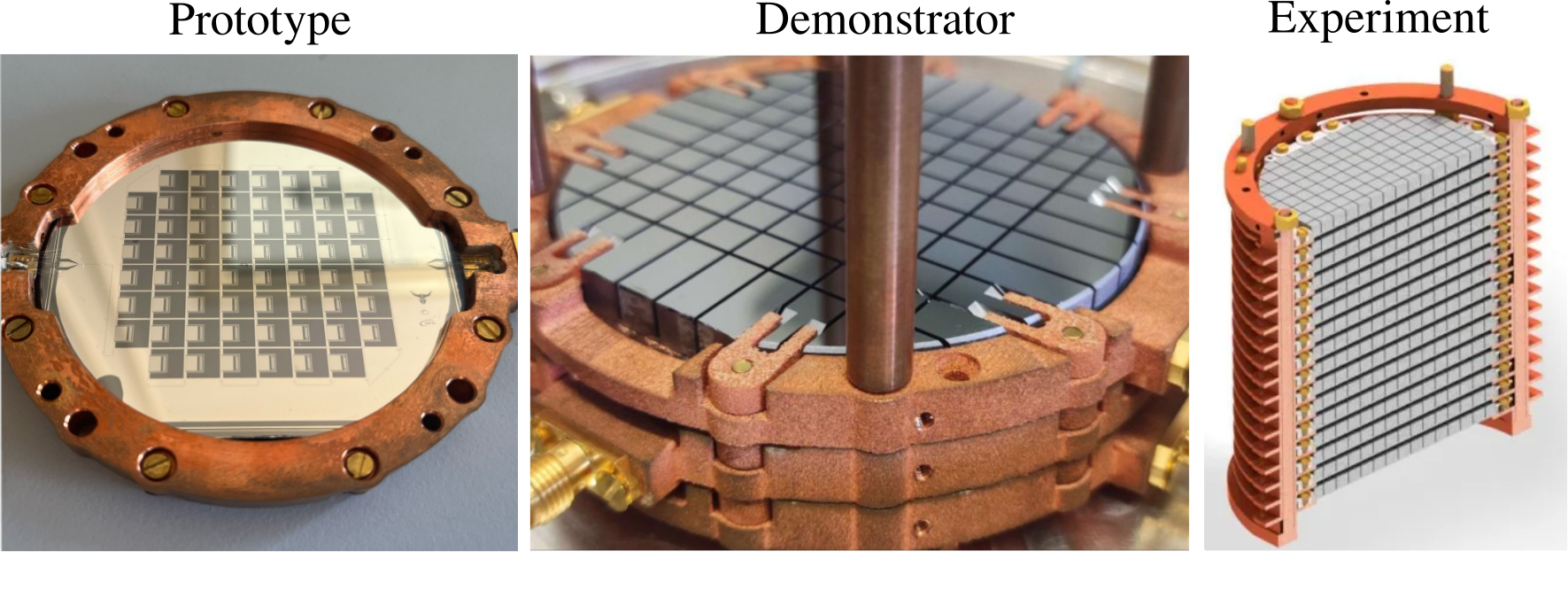}\\
    \centering
    \begin{tabular}{l|c|c|c|c}
      & 1. Prototype & \multicolumn{2}{c|}{2. Demonstrator} & 3. Experiment \\
     \hline
     \hline
     mass & 20 g & \multicolumn{2}{c|}{60 g} & 800 g\\
     \hline
     \# sensors & 60 & \multicolumn{2}{c|}{180} & 2300\\
     \hline
     Threshold & 160 eV & \multicolumn{2}{c|}{$\leq$ 200 eV} & $\leq$ 200 eV\\
     \hline
     Bkg rate (DRU) &  $ 2\cdot10^{6}$ & $< 10^{5}$ & $\sim 10^{4}$ & $1 \div 0.01$\\
     \hline
     Laboratory &  Sapienza U.&  Sapienza U. & LNGS & LNGS\\
     \hline
     Period  &  2023 & 2024 - 2026 & 2026 & 2027 - \\
     \hline
     \hline
    \end{tabular}
    \caption{Different stages of the BULLKID-DM road-map. From the left: the prototype, a $3''$ diameter, 5 mm thick silicon wafer; the demonstrator, consisting of a stack of 3 prototypes; the final detector design consisting of a stack of 16 wafers of 100 mm each. The table on the bottom summarizes the main properties of each stage of the road-map.}
    \label{fig:roadmap}
\end{figure}

BULLKID-DM is based on the technology developed by the BULLKID project. The BULLKID prototype ~\cite{Angelo_BULLKID,Delicato} consists of a $3''$ diameter, $5$ mm thick silicon wafer (Fig. \ref{fig:roadmap} left) carved on one side with $4.5$ mm deep grooves, creating dice capable of effectively containing a nuclear-recoil interaction. Simultaneously, this approach maintains a thin common disk shared among all the cubes, defining a modular system of a monolithic matrix of 60 dice with a total active mass of approximately $20$ g. 
When a particle interacts with the crystal, the generated a-thermal phonons are detected by an array of aluminum KIDs lithographed on the opposite side of the grooves, with one KID for each die. The design of the BULLKID-DM experiment consists on stacking $16$ BULLKID-like detectors of $100$ mm diameter and dice of equal dimensions (Fig. \ref{fig:roadmap} right) for a target mass of $800$ g. The road-map of BULLKID-DM is presented in the table of Fig. \ref{fig:roadmap}, showing the intermediate technological steps required to reach the final configuration of the experiment, expected at the beginning of 2027. Specifically, the goals of each step are summarized below:
\begin{enumerate}
\item Operation of the prototype at the Sapienza laboratories, concluded in 2023, demonstrated that the present detector can achieve an energy threshold of 160 eV \cite{Delicato}. Moreover, the simultaneous acquisition of multiple KIDs was proved;
\item[2a.] Operation of the demonstrator (stack of 3 detectors) in presence of a mildly shielded environment at the Sapienza laboratories by mid 2026. The aim is a background acquisition of the multi-sensor stack with a background rate below $10^5$ $\text{counts}~/~(\text{keV}\cdot\text{kg}\cdot\text{d})$ (DRU). A single-detector measurement has been already concluded and will be described in a forthcoming paper \cite{lead};
\item[2b.] Installation of the demonstrator at the cryo-platform of Laboratori Nazionali del Gran Sasso (LNGS), with a background rate around $10^4$ DRU, by the end of 2026. The main scope is to verify the level of the background contaminating the low energy spectrum and validate the Monte Carlo simulations;
\item[3.] Installation of the final experiment at the cryo-platform, with the target background rate below 1 DRU, possibly down to $10^{-2}$ DRU, and an energy threshold better or equal than $200$ eV, in 2027.
\end{enumerate}
Currently the collaboration is working on the second technological step and has proved that the same detector response is obtained for bulk and surface events \cite{Americium}. In parallel, efforts are ongoing to port the BULLKID technology to a germanium substrate \cite{Germanium}, so that the final experiment could employ both elements, providing a stronger validation in the case that a DM signal were observed \cite{CRESST_Edelweiss}.

The strategy of the collaboration to achieve the low background rate required relies on the combined use of shielding materials and veto systems. The shielding will include three layers of roman lead, high-putity copper and polyethilyene at room temperature surrounding the crtyostat, as well as cryogenic case of ultra-pure copper enclosing the detector and thermally anchored to the mixing-chamber plate of the cryostat. In addition, the active volume of the experiment will be capable of tagging events producing multiple hits within the detector itself, thereby rejecting interactions that are highly unlikely to originate from DM. The outer silicon dice in the active volume will also define the fiducial volume by identifying and excluding interactions occurring near the boundaries. Finally, the collaboration is developing a cryogenic veto \cite{Lari} made of a scintillating material (such as BGO or GAGG), read out by a light-sensitive KID detector \cite{CALDER}, to further enhance particle-background discrimination.

\section{Sensitivity projections to WIMP-like particles}
\begin{figure}
    \centering
    \includegraphics[width=1\linewidth]{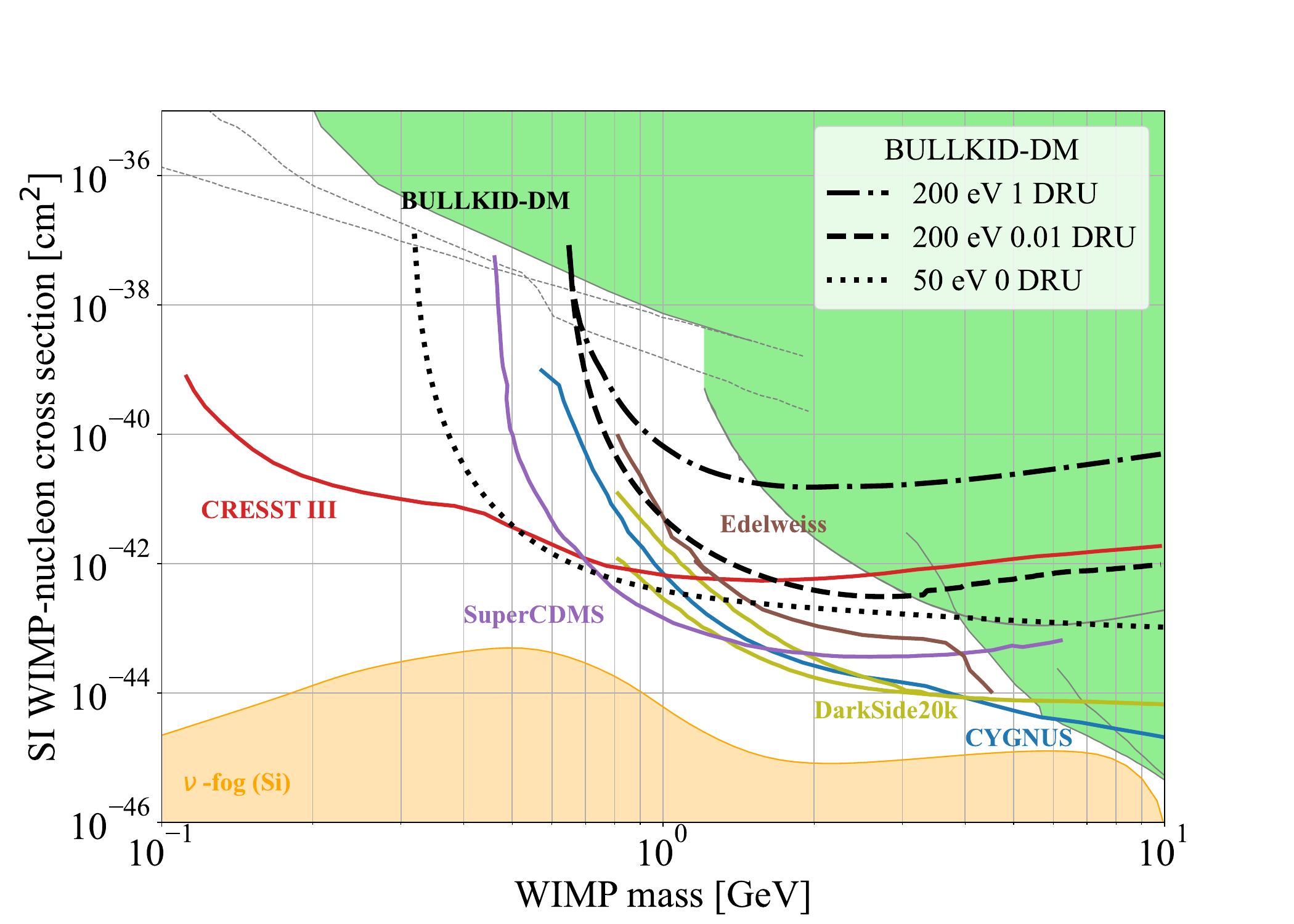}
    \caption{Projected sensitivities of BULLKID-DM and of other experiments (Tab. \ref{tab:future_exp}) as the 90 \% CL on the mass and the SI cross section of WIMPs with nucleons. The green area represents the state of the art; the orange region is the silicon neutrino fog \cite{nufog}.}
    \label{fig:state_of_art}
\end{figure}

To the best of our knowledge, the green area in Figure \ref{fig:state_of_art} summarizes the state of the art in direct WIMP detection as the 90\% Confidence Level (CL) of the SI cross-section of WIMPs with nucleons as a function of their mass. The black curves indicate our projections of the future impact of BULLKID-DM, for 1 year of exposure, $600$ g of fiducialized mass and for three different scenarios of energy threshold and background rate: the pessimistic case (200~eV, $1$~DRU), the optimistic case (200~eV, $10^{-2}$~DRU) and limit case (50~eV, 0~DRU). The projection are traced through the Yellin method \cite{Yellin}. The choice of this approach stems from its widespread adoption within the solid-state cryogenic detector community \cite{SuperCDMS-CPD,CRESST-III}, as it is capable of producing conservative yet robust limits even in presence of particle-background contributions that are not fully modeled or understood. In our case, at the cost of adopting a more conservative estimate of the future sensitivity, we applied the Yellin method to a simulated flat background in the three scenarios considered, thereby setting upper limits on the WIMP signal under the assumption of a completely unknown background. As the BULLKID-DM roadmap progresses, the experiment will gain an increasingly accurate understanding of the agreement between the measured and expected particle backgrounds, enabling a more refined projection of its sensitivity that incorporates explicit background modeling. The colored curves in Fig. \ref{fig:state_of_art} are the projected sensitivities of some other experiments in the field and Tab. \ref{tab:future_exp} summarizes their main properties. 
\begin{table}[b]
    \centering
    \begin{tabular}{l|c|c|c|c}
         Experiment & Target & Mass [kg] & Exposure [y] & Ref.  \\
         \hline
         \hline
         BULLKID-DM & Si & $0.8$ & 1 & This work \\
         DarkSide-20k & Ar & $49.7 \cdot 10^3$ & 1 & \cite{DarkSide-20k} \\
         CRESST III & CaWO$_4$ & $2.5$ & 1 & \cite{CRESST_Edelweiss} \\
         SuperCDMS & Ge & $11.12$ & 5 & \cite{Super-CDMS} \\
         Edelweiss & Ge & $20$ & 7 & \cite{CRESST_Edelweiss} \\
         CYGNUS & He-SF$_6$ & $167$ & 6 & \cite{Cygnus} \\
         \hline
         \hline
    \end{tabular}
    \caption{Summary of the main properties of the experiments for which the future projected sensitivities are traced in Fig. \ref{fig:state_of_art}. }
    \label{tab:future_exp}
\end{table}
\section{Conclusion}
BULLKID-DM is an experiment designed for the direct search of particle dark matter candidates with mass below 1 GeV. In this work we have introduced the detector concept and presented the expected road-map. The plan is to install the final experiment at the cryo-platform of LNGS in 2027, after previous technological steps with the aim of validating the performances of the detectors, the mass-scalability of the sensitive volume and the agreement between the measured and the simulated background from natural radioactivity. Finally we presented the future projected sensitivity on the SI cross-section of WIMPs with nucleons as a function of their mass. However, the collaboration is also working in parallel on R\&D activities with the aim of reducing the energy threshold \cite{Funnel} and, hence, increase the sensitivity to lower WIMP masses.

\acknowledgments
This work was further supported by the INFN, Sapienza University of Rome and co-funded by the European Union (ERC, DANAE, 101087663). Views and opinions expressed are however those of the author(s) only and do not necessarily reflect those of the European Union or the European Research Council. Neither the European Union nor the granting authority can be held responsible for them.

\end{document}